\documentclass[useAMS,usenatbib]{mn2e}
\usepackage{graphicx}

\title[Dynamical evolution of the Hygiea asteroid family]
{Dynamical evolution and chronology of the Hygiea asteroid family}
\author[V. Carruba, R. C. Domingos, M. E. Huaman, C. R. dos Santos, 
 and D. Souami]{V. 
Carruba$^{1}$\thanks{E-mail: vcarruba@feg.unesp.br}, R. C. Domingos$^{2}$,
M. E. Huaman$^{1}$, C. R. dos Santos$^{1}$, and D. Souami$^{3,4,5}$\\
$^{1}$UNESP, Univ. Estadual Paulista, Grupo de din\^{a}mica Orbital e
  Planetologia, Guaratinguet\'{a}, SP, 12516-410, Brazil \\
$^{2}$INPE, Instituto Nacional de Pesquisas Espaciais, S\~{a}o Jos\'{e} 
  dos Campos, SP, 12227-010, Brazil\\
$^{3}$NAXYS, Namur Center for Complex Systems, Department of Mathematics, 
University of Namur, 5000 Namur, Belgium\\
$^{4}$UPMC, Universit\'{e} Pierre et Marie Curie, 4 Place Jussieu, 75005, 
Paris, France\\
$^{5}$SYRTE, Observatoire de Paris, Syst\`{e}mes de R\'{e}f\'{e}rence 
Temps Espace, CNRS/UMR 8630, UPMC, Paris, France;
}

\begin{document}

\date{Accepted 2013 October 21.  Received 2013 October 18; in original form 
2013 August 20}

\pagerange{\pageref{firstpage}--\pageref{lastpage}} \pubyear{2013}

\maketitle

\label{firstpage}

\begin{abstract}
The asteroid (10) Hygiea is the fourth largest asteroid of the Main 
Belt, by volume and mass, and it is the largest member of its own family.
Previous works investigated the long-term effects of close encounters
with (10) Hygiea of asteroids in the orbital region of the family, and
analyzed the taxonomical and dynamical properties of members of this
family.  In this paper we apply the high-quality SDSS-MOC4 
taxonomic scheme of DeMeo and Carry (2013) to members of the Hygiea
family core and halo, we obtain an estimate of the minimum time
and number of encounter necessary to obtain a $3\sigma$ (or 99.7\%) 
compatible frequency distribution function of changes in proper 
$a$ caused by close 
encounters with (10) Hygiea, we study the behavior of asteroids near 
secular resonance configurations, in the presence and absence of
the Yarkovsky force, and obtain a first estimate of the age of the
family based on orbital diffusion by the Yarkovsky and YORP effects
with two methods.  

The Hygiea family is at least 2 Byr old, with an estimated age of 
$T = 3200^{+380}_{-120}$~Myr and a relatively large initial ejection
velocity field, according to the approach of Vokrouhlick\'{y} {\em et al.} 
(2006a, b). Surprisingly, we found that the family age can be shortened
by $\simeq$ 25\% if the dynamical mobility caused by close encounters
with (10) Hygiea is also accounted for, which opens interesting
new research lines for the dynamical evolution of families
associated with massive bodies.  In our taxonomical analysis
of the Hygiea asteroid family, we also identified a new V-type candidate: 
the asteroid (177904) (2005 SV5).  If confirmed, this
could be the fourth V-type object ever to be identified in 
the outer main belt.  

\end{abstract}

\begin{keywords}
Minor planets, asteroids: general -- Minor planets, asteroids: individual:
Hygiea -- Celestial mechanics.  
\end{keywords}
%

\section{Introduction}
\label{sec: intro}

The asteroid (10) Hygiea is the fourth largest asteroid of the Main 
Belt, by volume and mass, and it is the largest member of its own family.
The long-term effect of close encounters of asteroids in the orbital region
of the Hygiea family was recently investigated in Carruba et al. (2013a), 
that found surprisingly high values of drift rates of changes in proper
semi-major axis caused by this mechanism of dynamical mobility.
A preliminary taxonomical analysis and review of 
physical and dynamical properties of local asteroids was performed 
in Carruba (2013) (paper I hereafter), 
that found a somewhat limited role for secular dynamics
in the region of the Hygiea family, core and halo.
In this work we try to answer some of the questions posed by these two earlier
papers, and in particular we apply for the first time the high-quality 
taxonomy scheme described in the recently submitted paper of DeMeo and 
Carry (2013) to asteroids in the region, to eliminate all possible taxonomical
interlopers.   We extend the simulations carried out in Carruba et al. (2013a)
(paper II hereafter)
to get an estimate of the minimum number of close encounters needed
to obtain a $3\sigma-$level (or 99.7\%)  
approximation of the probability distribution 
function ($pdf$) of changes in proper $a$ caused by close encounters with 
(10) Hygiea.  We studied the actual behavior of the ``likely resonators''
identified in paper I to check for the fraction of objects currently
in resonant states, with and without the Yarkovsky force.  
We obtain for the first time an estimate of the age of the Hygiea family 
core and halo, with two independent methods, by considering the evolution
of asteroids semi-major axis under the Yarkovsky and YORP effects. Then,
we evaluate the effect that dynamical mobility caused by close encounters
with (10) Hygiea has on the estimated age of the family.

In this work we used values of asteroid proper elements, frequencies,
photometry, and albedos obtained from public databases.
The synthetic proper elements and frequencies are available at the 
AstDyS site 
http://hamilton.dm.unipi.it/cgi-bin/astdys/astibo, 
accessed on May $15^{th}$, 2013 (Kne\v{z}evi\'{c} and Milani 2003)
We used photometric data from the Sloan Digital Sky 
Survey-Moving Object Catalog data, fourth release (SDSS-MOC4 
hereafter, Ivezic et al. 2002), that provided multi-band photometry for
a sample two order of magnitude larger than any current available in 
spectroscopic catalogs (about 60000 numbered objects).
Finally, results from the Wide-field Infrared Survey Explorer (WISE) 
(Wright et al. 2010), and the NEOWISE (Mainzer et al. 2011) enhancement 
to the WISE mission recently allowed to obtain diameters and 
geometric albedo values for more than 100,000 Main Belt asteroids 
(Masiero et al. 2011), increasing the sample of objects for which
albedo values are known by a factor 50.  

This work is so divided:  in Sect.~\ref{sec: hygiea_det} we identified
members of the Hygiea family core and halo using the new taxonomical scheme
of DeMeo and Carry (2013).  In Sect.~\ref{sec: pdf} we studied the 
long-term effect of close encounters of asteroids in the orbital region
of (10) Hygiea with (10) Hygiea itself.  Sect.~\ref{sec: sec_dyn}
is decicated to the analysis of the dynamical evolution caused by secular
dynamics in the Hygiea family orbital region.  In Sect.~\ref{sec: cronology}
we estimated the age of the Hygiea family when evolution in semi-major
axis caused by Yarkovsky and YORP effects is considered.
In Sect.~\ref{sec: ce_chron} we considered how close encounters
affected the estimate of the family age.  Finally, in Sect.~\ref{sec: concl}
we present our conclusions.

\section{Hygiea family: identification and taxonomy}
\label{sec: hygiea_det}

The Hygiea family was most recently identified in the domain of proper elements
(a,e,sin(i)), in several domains of proper frequencies (Carruba 2013), 
and in a multi-domain of proper elements, SDSS-MOC4 (a*,i-z) colors, and WISE
geometrical albedo by Carruba et al. (2013b).  DeMeo and Carry (2013) recently
introduced a new classification method, based on the Bus-DeMeo taxonomic 
system, that employs  SDSS-MOC4 gri slope and z' -i' colors.
In that article the authors used the photometric data obtained
in the five filters $u', g', r', i'$, and $z'$, from 0.3 to 1.0 
$\mu m$, to obtain
values of  $z'-i'$ colors and spectral slopes over the $g', r'$, and $i'$ 
reflectance values, computed using the equation:

\begin{equation}
R_f = 10^{-0.4[(M_f-M_g)-(M_{f,sun}-M_{g,sun})]},
\label{eq: refl_sdss}
\end{equation}

\noindent where $M_f$ and $M_{f,sun}$ are the magnitudes of
the object and the sun in a certain filter f, respectively, 
at the central wavelength of that filter.  The equation is
normalized to unity at the central wavelength of filter
$g$ using $(M_g)$ and $M_{g,sun}$, the $g$ magnitudes of the 
object and the sun, respectively.  Values of the solar colors
$r'-g' = -0.45\pm 0.02, i'-g' = -0.55 \pm 0.03$, and 
$z'-g' = -0.61 \pm 0.04$ are taken from Holmberg et al. (2006). 

DeMeo and Carry (2013) defined strict criteria to reject flawed 
observation: they eliminated from the SDSS-MOC4 database
objects with a provisional designation, observations with 
unreliable magnitudes in any of the five filters, values
of the $u'$ filter (also because of the large errors associated
with measurements in this wavelength band), and data with flags 
relevant to moving objects and good photometry.  We refer the
reader to Sect. 2 of DeMeo and Carry (2013) for more details on 
the criteria used to obtain high quality measurements from
SDSS-MOC4 data.

We applied the methods described in DeMeo and Carry (2013) to the 
SDSS-MOC4 dataset.  We refer the reader to 
Fig.~5 in DeMeo and Carry (2013), that displays the 
boundaries, in the plane
$gri$ slope (measured in the standard units of $\%/100~nm$) 
versus $z'-i'$ colors, used to classify SDSS-MOC4 data into the 
taxonomic classes of the Bus-DeMeo taxonomy, converted to SDSS-MOC4 
gri slopes and $z'-i'$ colors.  We also assigned numbers
to the objects that, at the time of the release of SDSS-MOC4 had
only temporary designation and received numbers since then (a total of 7234 
asteroids), and, as in DeMeo and Carry (2013), we eliminated all objects 
with $H > 15.30$, so as to avoid including objects with $D < 5~$km, 
for which the sample is incomplete.  Also, we included asteroids 
with $H < 12.00$ with known spectral types from the Planetary Data 
System (Neese 2010) that are not part of SDSS-MOC4.

\begin{figure*}

  \centering
  \begin{minipage}[c]{0.5\textwidth}
    \centering \includegraphics[width=2.5in]{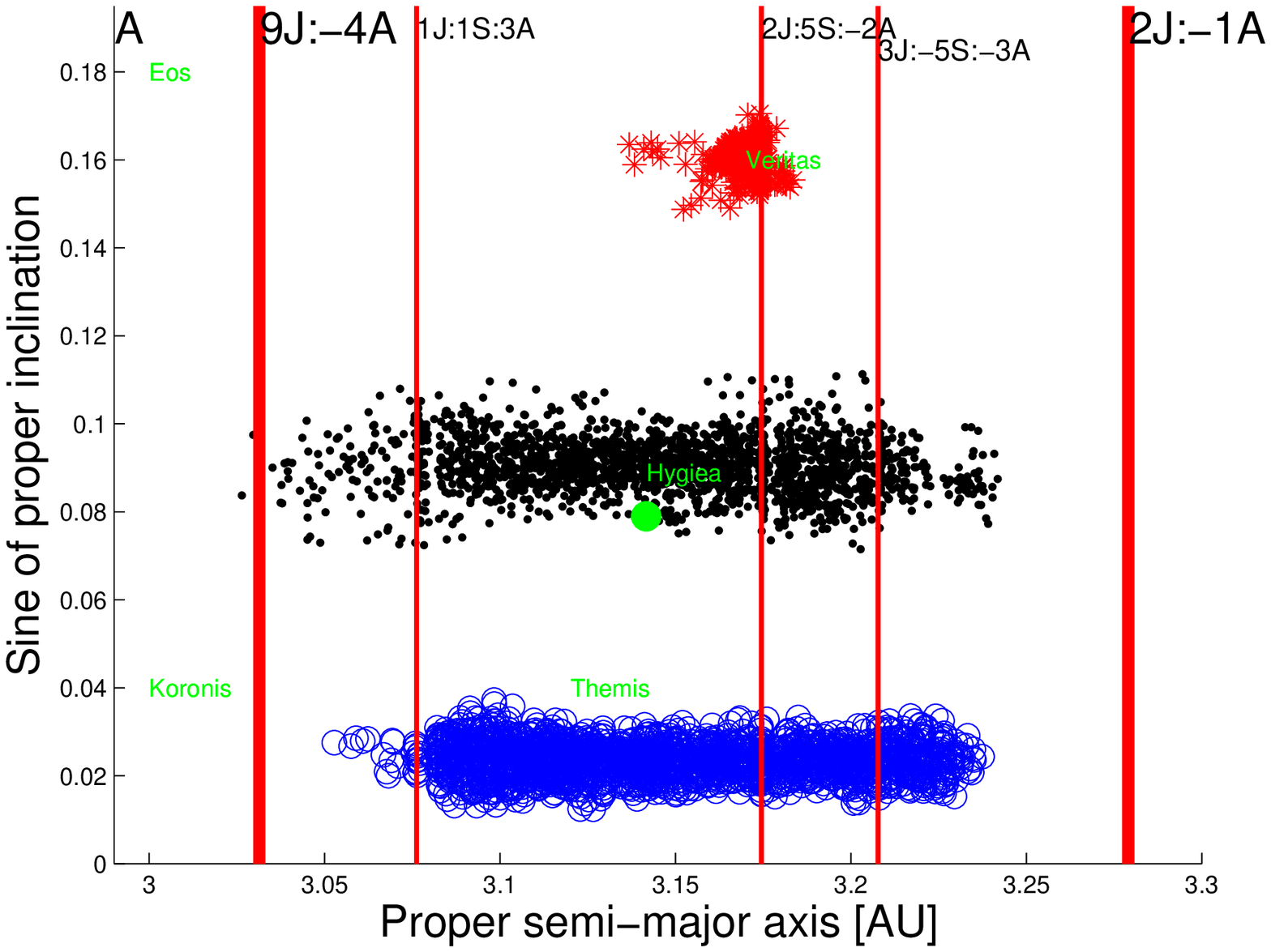}
  \end{minipage}%
  \begin{minipage}[c]{0.5\textwidth}
    \centering \includegraphics[width=2.5in]{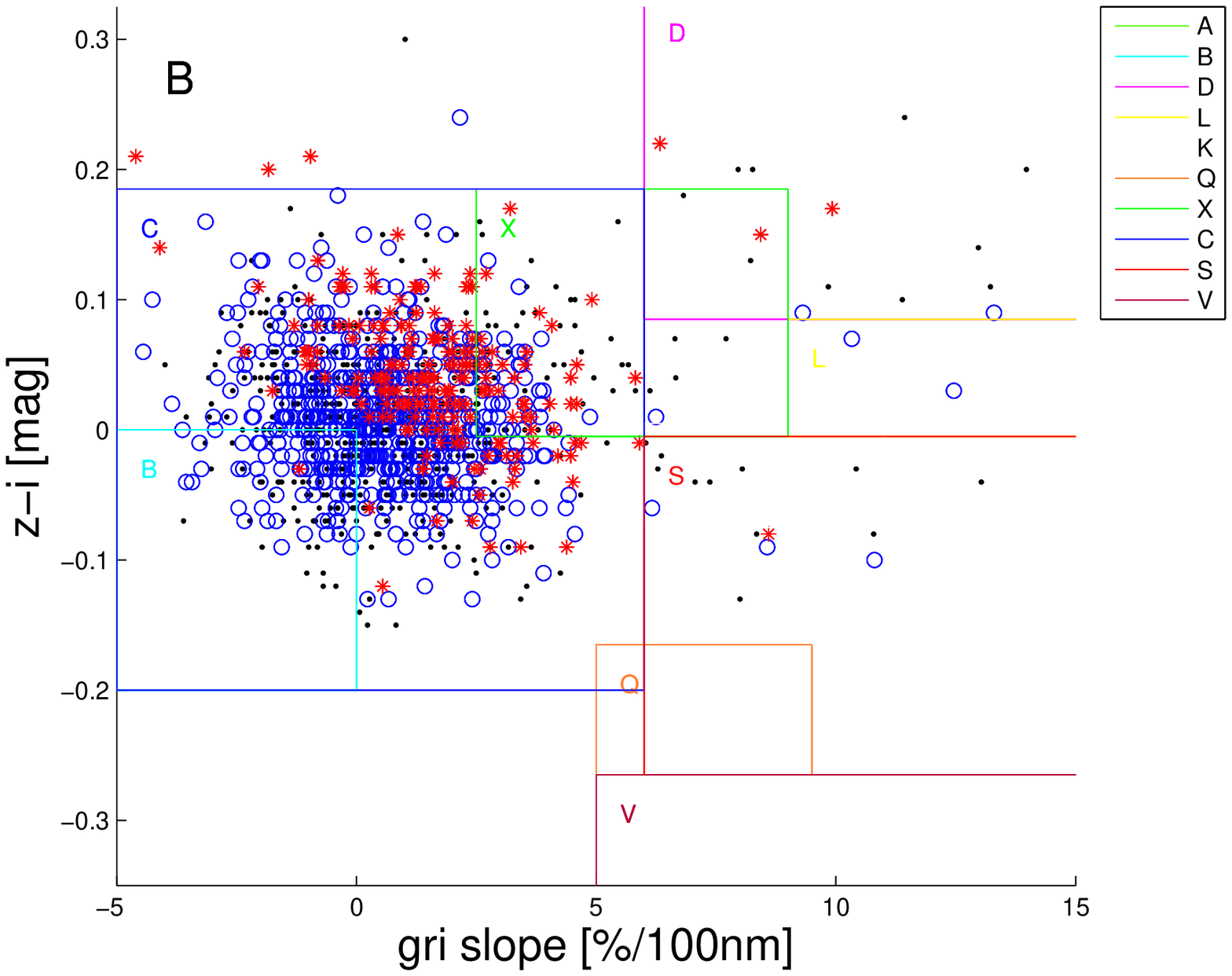}
  \end{minipage}

\caption{A proper $(a,sin(i))$ projection of members of 
the core of the Veritas (red asterisks), Hygiea (black dots), 
and Themis (blue circles) asteroid families, 
computed at a cutoff of 50 m/s in the proper element domain 
(panel A).  A projection of the same asteroids in the $gri$
slope versus $z'-i'$ plane, superimposed to the boundaries
of the DeMeo and Carry classification (panel B). 
}
\label{fig: fam_cores}
\end{figure*}

As a first part of our analysis, we checked if the preliminary 
analysis of the taxonomy of the Hygiea family halo perfomed
in paper I still holds.  In that article it was found that the 
composition of the Hygiea family as deduced from principal 
components obtained from SDSS-MOC4 photometry was very similar to that 
of the nearby Themis and Veritas families, all belonging to CX-complex 
taxonomies.  Therefore, it was somewhat difficult to determine, based only 
on taxonomical considerations, if a member of the Hygiea family halo (or 
indeed of the family core itself) was a fragment of the Hygiea parent 
body or a member of the two other families that dynamically migrated
to its present orbital location.  Here we applied the more advanced
taxonomical scheme of DeMeo and Carry (2013) to members of the cores
of the Hygiea, Themis, and Veritas families computed at a conservative 
cutoff value of $50$ m/s (or distance) in the proper element domain, 
and that also have data in 
the reduced SDSS-MOC4 sample.  Fig.~\ref{fig: fam_cores} displays 
proper $(a,sin(i))$ projection (panel A) and a $gri$
slope versus z'-i' plane projection, superimposed to the boundaries
of the DeMeo and Carry classification (panel B), of members of 
the core of the Veritas (red asterisks), Hygiea (black dots), 
and Themis (blue circles) asteroid families.

We believe that our results essentially confirm the preliminary analysis
of paper I:  most of the members of the Hygiea, 
Themis and Veritas families belong to the C- and X- type, with a fraction
of Hygiea and Themis family members that can be classified as B-types.
Essentially, C- and X- type members of the Hygiea family halo 
that may have come from the Themis or Veritas families are indistinguishable
from halo member that came from the parent body of (10) Hygiea itself.
B-type asteroids in that halo may either come from the Hygiea or Themis
families, but not from the Veritas one.  Eliminating possible Themis
or Veritas families interlopers in the Hygiea family halo based
on taxonomical considerations only seems, therefore, somewhat unlikely.

Since our focus in this paper is to study the Hygiea asteroid family, we 
selected objects in the SDSS-MOC4 reduced dataset that were
members of the Hygiea family core and halo, as obtained at velocities
cutoffs of 66 and 76 $m/s$ in paper I. We found a total of 
497 observations in the Hygiea family core and 695 observations in the Hygiea 
family halo.  A taxonomy is given according to the method described
in DeMeo and Carry (2013), i.e., if the $z'-i'$ and $gri$ slope is
found in the region boundary of a class then the asteroid is assigned
to that class.  For overlapping classes, the taxonomy is assigned 
in the last class in which the object resides, in the following
order: C-, B-, S-, L-, X-, D-, K-, Q-, V-, and A-types.
Since some asteroids had multiple observations in 
the SDSS-MOC4 database, we adopted the DeMeo and Carry (2013) criteria
for classifications in this cases:  in case of conflicts, 
the class with the majority number of classifications is assigned.  If two
classes have equal frequency, preference is given to C-, S-, or X-type 
classifications.  If the two majority classes are C/S, X/C or S/X, or there
is no majority, no class is assigned to this object  (class U in the 
DeMeo and Carry (2013) paper).  

\begin{table*}
\begin{center}
\caption{{\bf Number of C-, X-, B-, D-, K-, S-, L-
and U- type objects in the hygiea family core and halo.}}
\label{table: hygiea_tax}
\vspace{0.5cm}
\begin{tabular}{|c|c|c|c|c|c|c|c|c|}
\hline
            &     &     &    &    &   &    &   &   \\
Group &  \# of C-types & \# of X-types & \# of B-types & \# of D-types & 
\# of K-types & \# of S-types  & \# of L-types & \# of U-types \\
            &     &     &    &    &   &    &   &   \\
\hline
            &     &     &    &    &   &    &   &   \\
Hygiea core & 206 &  28 & 42 &  7 & 4 &  4 & 0 & 10\\
Hygiea halo & 270 &  56 & 50 & 19 & 4 & 10 & 6 & 16\\
            &     &     &    &    &   &    &   &   \\
\hline
\end{tabular}
\end{center}
\end{table*}

\begin{figure}
  \centering
  \centering \includegraphics [width=0.45\textwidth]{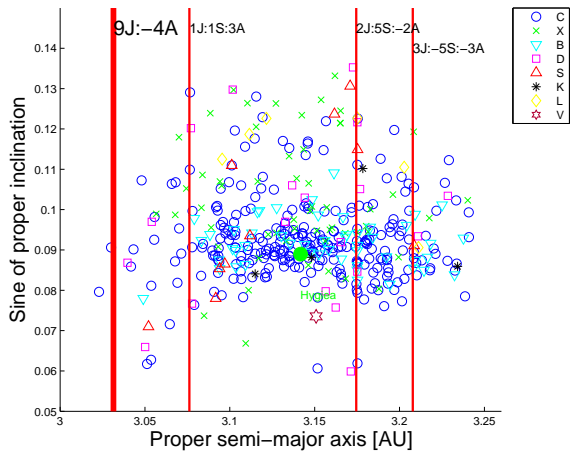}

\caption{A proper $(a,sin(i))$ projection of asteroids in the Hygiea family
halo with an identifiable SDSS-MOC4 taxonomy, according to the DeMeo 
and Carry (2013) scheme.} 
\label{fig: hygiea_tax_sdss_ai}
\end{figure}

Table~\ref{table: hygiea_tax} displays the number of C-, X-, B-, D-, K-, S-, 
L-, and U- type objects in the hygiea family core and halo, whose
orbital projection in the plane of proper $(a,sin(i))$ is given 
in Fig.~\ref{fig: hygiea_tax_sdss_ai}.  We also, 
very surprisingly, identified one possible V-type object, 
the asteroid 177904 (2005 SV5), that, to our knowledge, could be the fourth
V-type object found in the outer main belt 
after the well known case of (1459) Magnya, and (7472) Kumakiri
and (10537) (1991 RY16) (Duffard 2009).
According to the mineralogic analysis of Moth\'{e}-Diniz et al. (2005), 
D-, S-, K-, V-, A- 
and U- type objects are to be considered interlopers for taxonomical reasons
and will not be considered part of the family, made mostly 
by B-, C-, and X-type objects, for what concerns its
age determination.  This analysis left us with 276 and 376
$D > 5$~km asteroids in the Hygiea family core and halo, respectively.
The mean WISE albedo of the Hygiea family core members was of 
0.0576, with a minimum value of 0.0245 and a maximum of 0.189
~\footnote{Since WISE albedo values tend to be higher
than usually expected for small objects, we eliminated from our
sample four asteroids with geometric albedos higher that 0.2, a value
this not usually associated with C-, X-, and B- type asteroids.
Including these asteroids in our sample yields the sligthly higher
value of mean WISE albedo of 0.078.}.  These
results are compatible with what observed for typical C- and B- type
classes in DeMeo and Carry (2013).

Having obtained a determination of the orbital boundaries of the Hygiea
family, we are now ready to start studying orbital dispersion mechanisms
such as the long-term effect of close encounters with (10) Hygiea.  This 
will be the subject of the next section.

\section{Long-term effect of close encounters with 
(10) Hygiea: Probability distribution function}
\label{sec: pdf}

The long-term effect of close encounters of asteroids in the orbital region
of (10) Hygiea with (10) Hygiea itself was most recently studied by paper 
II.  In that article, the authors found that the frequency 
distribution functions ($fdf$) 
of changes in proper $a$ caused by close encounters
tested positive against the null hypothesis at a $2\sigma$-level, 
or 95.4\% probability level, 
for number of encounters larger than 4500, and were expected
to be compatible amongst them at a $3\sigma$-level, or 99.7\% probability
level, for number of encounters larger than $\simeq 6000$, based on 
analytical considerations obtained using Greenberg (1982) 
model of the effect of changes in heliocentric velocities caused
by close encounters.  Essentially, the higher the number of encounters, the
better the phase space of minimum distance and relative velocity at 
encounter, the two parameters that dictate the change in heliocentric
velocity of the asteroid, and, therefore, of proper elements, the more
precise the approximation of the $fdf$ to the real probability distribution
function (or $pdf$).  The actual minimum number of encounters needed
to obtain a $3\sigma$-level approximation of the $pdf$ was, however, not
tested in that work.

\begin{figure}
  \centering
  \centering \includegraphics [width=0.45\textwidth]{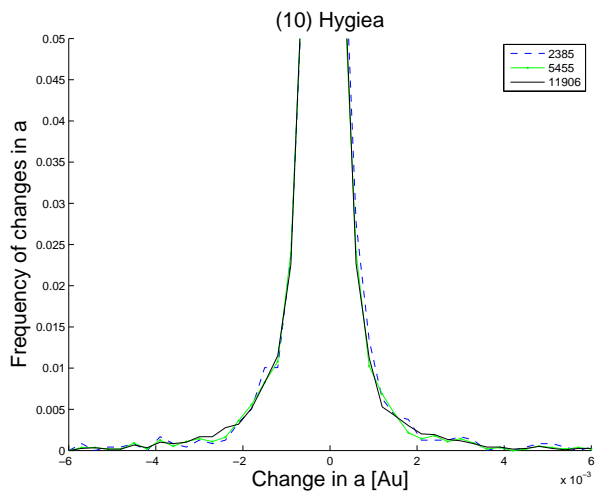}

\caption{Histogram of frequency of changes in proper $a$ caused by 
close encounters with (10) Hygiea.} 
\label{fig: hist_da}
\end{figure}

To verify the actual number of close encounters needed to obtain
a $3\sigma$-level approximation of the $pdf$, we performed a simulation with 
7015 particles belonging to the Hygiea family halo as identified in 
paper I.  Since paper II showed that, for the purpose
of obtaining a good estimate of the $pdf$ results obtained also 
including the Yarkovsky effect are not quantitatively different from
those obtained with simulations without such effect, we limited
our modeling to the effect of close encounters with (10) Hygiea.
The 7015 particles were integrated under the influence of
the eigth planets and the four most massive bodies in 
the main belt, (1) Ceres, (2) Pallas, (4) Vesta, and 
(10) Hygiea, over 34 Myr using the SWIFT-SKEEL 
code of the SyMBA package of Levison and Duncan (1994), modified
to monitor each encounter that occurred between (10) Hygiea and 
the test particle at a distance of less than $1\cdot 10^{-3}~AU$ 
and with a time step of 2 days (see Carruba et al. (2012) for a discussion
on the reasons for choosing these two parameters).
Since the standard deviation of changes obtained in the absence
of encounters with (10) Hygiea is of the order of $2 \cdot 10^{-4}$~AU,
we concentrated our attention on encounters that caused a change
in $a$ at least three times this value, since these are more significant 
for the dynamical mobility caused by close encounters with massive 
asteroids (Carruba et al. 2003), and are less likely to be 
caused by other effects.

Fig.~\ref{fig: hist_da} displays a histogram of frequency of changes 
in proper $a$ caused by close encounters with (10) Hygiea for 
our simulated particles.  As in paper II we performed
Kolmogorod-Smirnoff probability tests (KS tests hereafter) for each of the 
observed distributions at confidence levels of $2\sigma$ and $3\sigma$
between the whole distribution of 11906 encounters occurred during 
the simulation, and sub-samples of encounters, 
in the regions of changes in $a$ of most interest, i.e., 
$0.0006 < |\Delta a| < 0.006$~AU.  We found that sub-samples
of encounters start to be compatible at $2\sigma$ level for
2385 encounters, and at a $3\sigma$ level for 5455 number of encounters,
that is in remarkable good agreement with what predicted in paper II.  
The frequency distribution functions of $\Delta a$
for the 2385 and 5455 encounters are also displayed in 
Fig.~\ref{fig: hist_da}.

In Sect.~\ref{sec: hygiea_det} we identified 276 and 376 taxonomically 
compatible asteroids in the Hygiea family core and halo, respectively.
How much time would be necessary for these populations of asteoroids
to experience number of encounters such that the $fdf$ in $\Delta a$
could converge to the $pdf$, at least at a $2\sigma$ confidence level?
To answer this question, we computed the number of encounter experienced
by these two populations during the 34 Myr integration,
as a function of time, and extrapolated these values with simple linear laws.

\begin{figure}
  \centering
  \centering \includegraphics [width=0.45\textwidth]{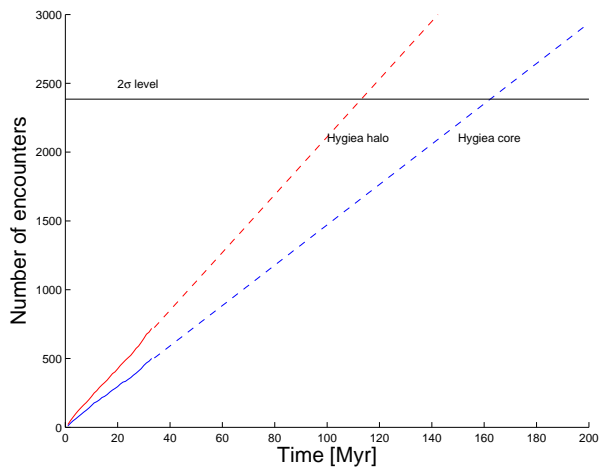}

\caption{Number (line) and expected number (dotted line) 
of encounters with (10) Hygiea for the population of objects with $H < 15.3$
in the Hygiea family core (blue lines) and halo (red lines).} 
\label{fig: proj_numb_enc}
\end{figure}

Fig.~\ref{fig: proj_numb_enc} displays the number (line) and expected 
number (dotted line) of encounters with (10) Hygiea for the population 
of objects with $H < 15.3$ in the Hygiea family core (blue lines) and 
halo (red lines).  To reach a number of 2385 encounters, sufficient
to attain a $2\sigma$ level in the approximation of the $pdf$, it is expected
that it will take 113.1 Myr for the halo population and 162.3 Myr 
for the core.  We will further investigate the importance of these
time-scales in Sect.~\ref{sec: cronology}.  In the next section we 
will bring our attention to the importance of secular dynamics in 
the Hygiea region.

\section{Secular dynamics in the Hygiea region}
\label{sec: sec_dyn}

A preliminary analysis of secular dynamics in the region of the Hygiea
family was already performed in paper I.  In that work the author
identified the population of likely resonators, i.e., the asteroids
within $\pm 0.3~arcsec/yr$ from the resonance center 
(for example, in the case of the $z_1$ secular resonance, a $g+s$ 
kind of resonance in the notation of Machuca and Carruba (2011), likely 
resonators would be objects whose combination of $g+s$ would fall 
to within ${\Delta}_{gs} = 0.3~arcsec/yr$ from the resonance center, 
i.e. $g+s = g_6+s_6  \pm {\Delta}_{gs} $).  Limitations of that 
approach are however that i) the $0.3~arcsec/yr$ limit was derived
for the $z_1$ resonance, and may differ for other secular resonances,
ii) the fact that an asteroid is a likely resonator does
not guarantee that it is in a librating state, and iii) the method
used to obtain synthetic proper elements may produce values
of proper frequencies close to resonant values, even for asteroids
in circulating states.  In order to
obtain a better understanding of secular dynamics in the region, we integrated
all the population of likely resonators identified in paper I
over 10 Myr with a Burlisch-Stoer integrator from the SWIFT package (Levison and
Duncan, 1994) modified by Bro\v{z} (1999) so as to include on-line
digital filtering to remove all frequencies with period less than 600
yr, under the influence of the eight planets.  We then obtained
the resonant argument of each resonance discussed in paper I, and
verified which asteroids were actually in librating states.
Results are summarized in Table~\ref{table: sec_res}.  With respect to
what found in paper I we notice that the population of 
actual resonators in secular resonances that cross the Hygiea family is
indeed quite limited, with the largest population of just six 
objects in states that alternate between libration and circulation 
of the ${\nu}_5+2{\nu}_{16}$ secular resonance.  No asteroids in pure
librating states were identified for this resonance.
The population
of librators is actually present in larger numbers for resonances that 
cross the Eos family orbital region, such as the $3{\nu}_6-2{\nu}_5$
(16 objects), the $2{\nu}_5-2{\nu}_6+{\nu}_{16}$ (69 librators), and
the ${\nu}_6+{\nu}_{16}$ (66 objects) secular resonances.  To investigate
the width of each of the major secular resonances observed in 
the region (for our purposes those with more than five observed librators) 
we used a resonant representative plane defined, 
as in Vokrouhlick\'{y} et al. (2006b) and in Carruba (2009a) by 
the critical angle $\sigma$ 
of each resonance and its conjugated frequency 
$d{\sigma}/dt (= g+s-g_6+s_6$ for the ${\nu}_6+{\nu}_{16}$ resonance), 
computed considering the 
effect of gravitational forces only.   We proceed as follows 
to compute these quantities: 
i) the orbital elements, results of the numerical simulation, 
are used to obtain equinoctial, non-singular elements of the form
$(e\cdot \cos{\varpi}, e\cdot \sin{\varpi})$, and
$(\sin{(i/2)}\cos{\Omega}, \sin{(i/2)}\sin{\Omega})$;
ii) the equinoctial elements of the test particles
and of Jupiter, Saturn, and Uranus, are then Fourier 
filtered to obtain the $g,s,g_5, g_6,s_6$
frequencies and their associated phases (see also Carruba 2010b for
details on the method);  iii) the frequencies are then
plotted on the ordinates and their phases are used to
construct the resonant angle $\sigma$. 

\begin{table*}
\begin{center}
\caption{{\bf Main secular resonances in the Hygiea region, frequency
    value, number of resonant asteroids, and limit on the resonance width}}
\label{table: sec_res}
\vspace{0.5cm}
\begin{tabular}{|c|c|c|c|}
\hline
                 &                   &                              & \\
Resonance argument & Frequency value & Number of resonant asteroids & Resonance width  \\
                 &  $arcsec/yr$  &          &       $arcsec/yr$  \\
\hline
                 &                   &          &                     \\
                 &  g resonances     &          &                    \\
                 &                   &          &                    \\
$2{\nu}_6-2{\nu}_5+{\nu}_7$& 51.065  &     4     &                   \\
$3{\nu}_6-2{\nu}_5$&  76.215        &     16     &  $< 0.188$        \\
                 &                   &          &                   \\
                 & s resonances      &          &                    \\
                 &                   &          &                   \\
$2{\nu}_5-2{\nu}_6+{\nu}_{16}$& -74.317 &    69   &  $<0.215$         \\  
                 &                   &          &                    \\
                 & g+s resonances    &          &                   \\
                 &                   &          &                    \\
$2{\nu}_5-{\nu}_6+{\nu}_{16}$ & -46.074 &    1    &                    \\
${\nu}_5+{\nu}_{16}$&  -22.088        &      2    &                    \\
${\nu}_6+{\nu}_{16}$&  1.898          &     66    &  $<0.331$         \\      
$2{\nu}_6-{\nu}_5+{\nu}_{16}$& 25.884 &      5    &                   \\
$2{\nu}_6-{\nu}_5+{\nu}_{17}$& 49.233 &      3    &                   \\
                 &                   &           &                   \\
                 & g+2s resonances   &        &                  \\
                 &                   &           &                   \\
${\nu}_5+2{\nu}_{16}$& -48.433        &      6    & $< 0.127$          \\
${\nu}_6+2{\nu}_{16}$& -24.447        &      4    &                   \\
                 &                   &           &                   \\
                 & 2g+s resonances   &        &                   \\
                 &                   &           &                   \\
${\nu}_5+{\nu}_6+{\nu}_{16}$& 6.155   &    1      &                   \\
                 &                   &          &                    \\
\hline
\end{tabular}
\end{center}
\end{table*}

\begin{figure}

  \centering
  \centering \includegraphics [width=0.45\textwidth]{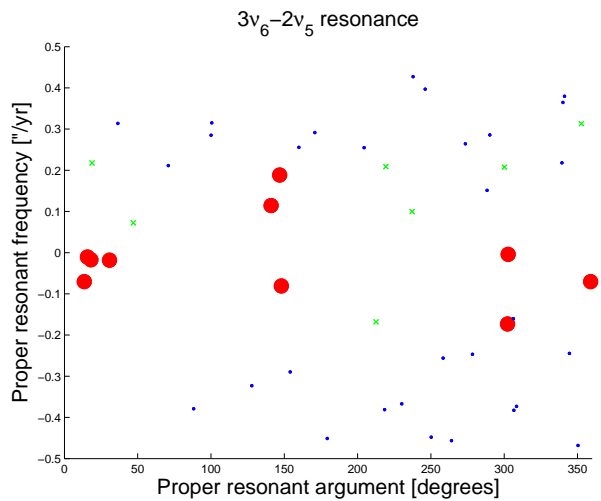}

\caption{A projection in the plane $(\varpi+2 \cdot {\varpi}_5-3 
\cdot {\varpi}_6, g+2g_5-3g_6)$ of asteroids in librating states (red full
dots), circulating states (blue dots), and alternating phases of librations
and circulations (green asterisks).} 
\label{fig: repr_3nu6-2nu5}
\end{figure}

Fig.~\ref{fig: repr_3nu6-2nu5} displays our results for the $3{\nu}_6-2{\nu}_5$
secular resonance, where asteroids in librating states are shown as red full
dots, those in circulating states are marked as blue dots, and 
the asteroids alternating phases of librations
and circulations are displayed as green asterisks.  We found that the 
width of the region populated by librating asteroids is defined by 
$g = (3g_6-2g_5)^{+0.188 ``/yr}_{-0.173 ``/yr}$, i.e., less than the 
$0.3~arcsec/yr$ criteria used for the $z_1$ resonance.  

\begin{figure}
  \centering
  \centering \includegraphics [width=0.45\textwidth]{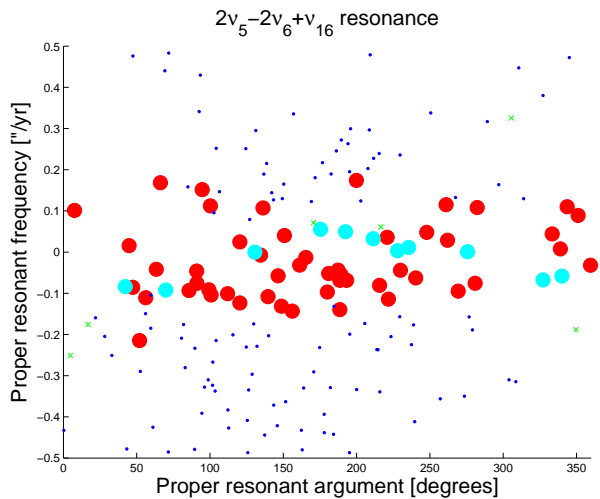}

\caption{A projection in the plane $(\Omega -{\Omega}_6+2 \cdot {\varpi}_6-2 
\cdot {\varpi}_5, s-s_6+2g_6-2g_5)$ of asteroids in librating states (red full
dots), librating states around $270^{\circ}$ (cyan full dots), 
circulating states (blue dots), and alternating phases of libration
and circulations (green asterisks).} 
\label{fig: repr_2nu5-2nu6+nu16}
\end{figure}

Fig.~\ref{fig: repr_2nu5-2nu6+nu16} displays the resonance representative
plane for the $2{\nu}_5-2{\nu}_6+{\nu}_{16}$
secular resonance, where we also identified a new class
of librating objects, oscillating around $270^{\circ}$. The 
width of the region populated by librating asteroids is defined by 
$g = (s_6+2g_5-2g_6)^{+0.174 ``/yr}_{-0.215 ``/yr}$, once again, less than the 
$0.3~arcsec/yr$ criteria used for the $z_1$ resonance.  

\begin{figure}
  \centering
  \centering \includegraphics [width=0.45\textwidth]{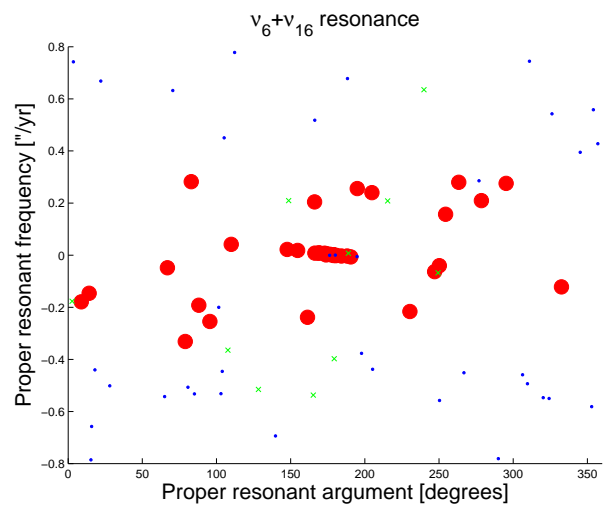}

\caption{A projection in the plane $({\varpi}-{\varpi}_6+ 
\Omega -{\Omega}_6, g-g_6+s-s_6)$ of asteroids in librating states (red full
dots), circulating states (blue dots), and alternating phases of libration
and circulations (green asterisks).} 
\label{fig: repr_nu6+nu16}
\end{figure}

Fig.~\ref{fig: repr_nu6+nu16} displays the resonance representative
plane for the ${\nu}_6+{\nu}_{16}$ secular resonance.  Compatibly with 
what was found in Carruba (2009a), the 
width of the region populated by librating asteroids is defined by 
$g = (g_6+s_6)^{+0.331 ``/yr}_{-0.282 ``/yr}$, i.e., in agreement with the
$0.3~arcsec/yr$ criteria used in paper I.

A similar analysis performed for the ${\nu}_5+2{\nu}_{16}$ showed
that i) there are no asteroids in pure librating states, and ii) the 
asteroids that had phases of libration during the simulation length and
that were the closest to the resonance center had a difference in resonant 
frequency of $1.27~arcsec/yr$, which sets an upper limit
on the resonance width.  In the next sub-section we will investigate
what fraction of the resonant population, that we identified, remains
in the resonances when non-gravitational forces such as the
Yarkovsky effect are considered.


\subsection{Yarkovsky evolution}
\label{sec: yarko_sec}

To investigate how effective the analyzed secular resonances are as
a mechanism of dynamical mobility, we integrated the real asteroids
that satisfied the preliminary frequency criterion in paper I
with SWIFT-RMVSY, the symplectic integrator of Bro\v{z} (1999) that 
simulates the diurnal and seasonal versions of the Yarkovsky effect, 
over 30 Myr and the gravitational influence of all planets from Venus 
to Neptune (Mercury was accounted for as a barycentric correction in 
the initial conditions). Since asteroids in the region are
mostly C-type, we used values of the Yarkovsky parameters appropriate 
for such bodies (Carruba et al. 2003): a thermal conductivity 
$K = 0.001 W/m/K$, a thermal capacity $C = 680~J/kg/K$, surface density 
1500 $kg/m^3$), a Bond albedo of 0.1, a thermal emissivity of 0.95, and a 
bulk density of 1500 $kg/m^3$.  We used two sets of spin axis orientations 
with $\pm90^{\circ}$ with respect to the orbital plane, since our goal is 
to investigate the maximum possible diffusion of asteroids and these 
obliquities maximize the speed of the Yarkovsky effect.  We assumed 
periods obtained under the approximation that the rotation frequency is
inversely proportional to the object's radius, and that a 1 km
asteroid had a rotation period of 5 hours (Farinella et al. 1998)
~\footnote{Other choices of rotation periods are possible.
One can choose a distribution of rotation frequencies similar to that
of other families, and randomly choose values for each asteroid.
However, Cotto-Figueroa et al. (2013) have shown that the 
YORP effect is extremely sensitive to the topography of the asteroid, 
and its small changes.  We therefore believe that 
in the end it may make little difference what initial rotation period 
is chosen.  The rotation period of a asteroid will change during a
YORP cycle in ways that are not currently well understood.
Since our goal in this section is to preliminary
investigate the fraction of surviving resonators when non-gravitational
forces are considered, we believe that our simpler approach was  
justified.}.
No re-orientations were considered, so that the drift caused by the
Yarkovsky effect was the maximum possible. 
We also used the WISE radii for the 295 asteroids for which this information 
was available, for the other objects we computed the radius using 
the equation:

\begin{equation}
R(km)= 664.5 \frac{10^{(-H/5)}}{\sqrt{p_V}},
\label{eq: bowell}
\end{equation}

\noindent
where $H$ is the asteroid's absolute magnitude provided by the AstDyS site, and
$p_V$ is the geometric albedo, assumed equal to that of (10) Hygiea, 
as measured from the WISE mission
($p_V = 0.0579$).  We computed proper elements with the approach described
in the previous section over 23 intervals of 1.2288 Myr (i,e. 2048
intervals of 600 yr, where 600 is the time interval in the output
of our simulation, and $2048 = 2^{11}$ is a power of 2 in order
to perform a Fourier analysis), and checked
what objects remained inside the investigated secular resonances, for
how long, and what was the change in proper $a,e,sin(i)$ caused by the 
passage through resonance.  We focused on resonances that the analysis
of paper I showed more likely to affect the evolution of the 
Hygiea asteroid family, such as the ${\nu}_5+2{\nu}_{16}$,
${\nu}_6+2{\nu}_{16}$, $2{\nu}_6-{\nu}_5+{\nu}_{16}$, and 
$2{\nu}_6-{\nu}_5+{\nu}_{17}$.   The $3{\nu}_6-2{\nu}_5$, 
$2{\nu}_5-2{\nu}_6+{\nu}_{16}$, and ${\nu}_6+{\nu}_{16}$ secular
resonances were also studied because the analysis of 
Sect.~\ref{sec: sec_dyn} showed us that they have the largest 
population of actual librators in the region, despite the 
fact that they only marginally affect the evolution of
Hygiea family asteroids.

\begin{table*}
\begin{center}
\caption{{\bf Number of librators, mean and maximal value of time of 
permanence in resonance, change in proper $e$ and $sin(i)$, for the studied 
resonances}}
\label{table: sec_res_analysis}
\vspace{0.5cm}
\begin{tabular}{|c|c|c|c|c|c|c|c|}
\hline
  &      &       &        &      &     &    &\\
Resonance argument & Number of librators & mean $\Delta T$ & max $\Delta T$ &
mean $\Delta e$ & max $\Delta e$ & mean $\Delta sin(i)$ & max $\Delta sin(i)$ \\
  &      &   [Myr] &  [Myr] &    &     &    &   \\
  &      &       &        &      &     &    &\\
\hline
  &      &       &        &      &     &    &\\
$3{\nu}_6-2{\nu}_5$           & 45& 21.4 & 30 & 0.030 & 0.050 & 0.0041 & 0.0160\\
$2{\nu}_5-2{\nu}_6+{\nu}_{16}$ & 96& 22.6 & 30 & 0.008 & 0.070 & 0.0055 & 0.0900\\
${\nu}_6+{\nu}_{16}$           &159& 22.9 & 30 & 0.024 & 0.170 & 0.0105 & 0.0520\\
${\nu}_5+2{\nu}_{16}$          & 15& 13.8 & 30 & 0.024 & 0.080 & 0.0050 & 0.0013\\
${\nu}_6+2{\nu}_{16}$          & 4 & 4.75 & 10 & 0.055 & 0.110 & 0.0002 & 0.0005\\
$2{\nu}_6-{\nu}_5+{\nu}_{16}$  & 29&16.17 & 30 & 0.007 & 0.035 & 0.0047 & 0.0400\\
$2{\nu}_6-{\nu}_5+{\nu}_{17}$  & 12& 10.2 & 30 & 0.005 & 0.010 & 0.0004 & 0.0010\\
  &      &       &        &      &     &    &\\
\hline
\end{tabular}
\end{center}
\end{table*}

Table~\ref{table: sec_res_analysis}, which reports the number of librators, 
mean and maximal value of time of permanence in resonance, changes in 
proper $e$ and $sin(i)$, for the studied resonances, summarizes our results.
Secular resonances in the Hygiea family area have a limited effect on asteroid
dynamics: the largest population of resonators was found inside the 
$2{\nu}_6-{\nu}_5+{\nu}_{16}$ resonance, with 29 asteroids.  Resonances in the 
Eos family area seem to have more impact on the local dynamics, which 
is especially true for the powerful ${\nu}_6+{\nu}_{16}$  secular resonance,
whose effect on the Eos family was studied in Vokrouhlick\'{y} et al. 
(2006d) and Carruba and Michtchenko (2007), among others.  While the 
population of resonators is limited in secular resonances in the Hygiea 
family region, the effect of such resonances is not negligible.  Maximum 
changes in proper excentricity
and inclination occur for particles whose orbits cross the separatrix between
circulation and libration\footnote{It should also be pointed out
that many particles in resonances such as the $3{\nu}_6-2{\nu}_5$ and 
the $2{\nu}_5-2{\nu}_6+{\nu}_{16}$ oscillate around more than one point of 
equilibrium, around $0^{\circ}$ and $180^{\circ}$.}.  The relatively 
weak ${\nu}_6+2{\nu}_{16}$ secular
resonance, in which none of the particles remained inside the resonance
for the whole length of the integration, caused some of the largest changes
in proper $e$ and $sin{(i)}$ observed when the particles switched from 
circulation
to libration.  The large number of weak secular resonances present in the
region may, in principle, provide an additional mechanism of mobility in 
$e$ and $sin{(i)}$ not considered in previous works and that may account
for some of the differences between the dispersion in proper $e$ and 
$sin{(i)}$ of observed and simulated families.  

\begin{figure}
  \centering
  \centering \includegraphics [width=0.45\textwidth]{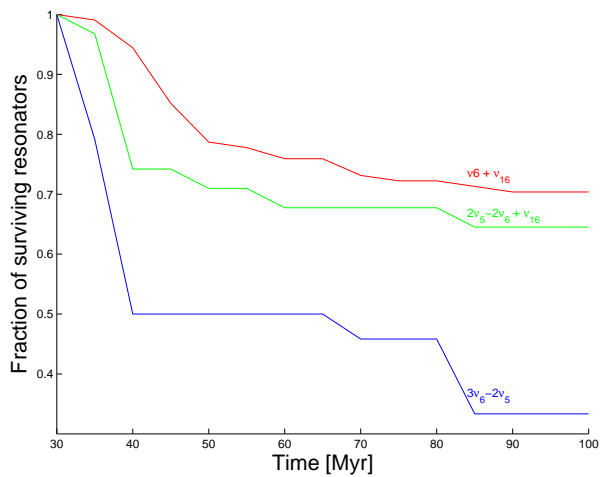}

\caption{The fraction of surviving resonators inside the ${\nu}_6+{\nu}_{16}$ 
(red line), $2{\nu}_5-2{\nu}_6+{\nu}_{16}$ (green line), and  
$3{\nu}_6-2{\nu}_5$ (blue line) secular
resonances as a function of time.} 
\label{fig: fract_surv}
\end{figure}

To better understand the long-term importance of evolution inside 
secular resonances
we also performed longer simulations for the resonators inside the three
secular resonances in the area with the largest population of librators: 
the $3{\nu}_6-2{\nu}_5$, $2{\nu}_5-2{\nu}_6+{\nu}_{16}$, and 
${\nu}_6+{\nu}_{16}$ resonances.  We computed the ``sticking time'', defined
as the time or permanence of particles
inside each resonance, and the changes in 
proper $e$ and $sin{(i)}$ associated with the passage through such 
commensurabilities.  Fig~\ref{fig: fract_surv}
displays the fraction of surviving resonators as 
a function of sticking times for the three studied resonances:  the longest 
times were observed for the  ${\nu}_6+{\nu}_{16}$, that also had the 
largest initial population of resonators (108, with respect to the 
31 and 24 of the $2{\nu}_5-2{\nu}_6+{\nu}_{16}$ and $3{\nu}_6-2{\nu}_5$
resonances, respectively).  Other studied resonances had lower numbers
of resonators and shorter sticking times.

\begin{figure}
  \centering
  \centering \includegraphics [width=0.45\textwidth]{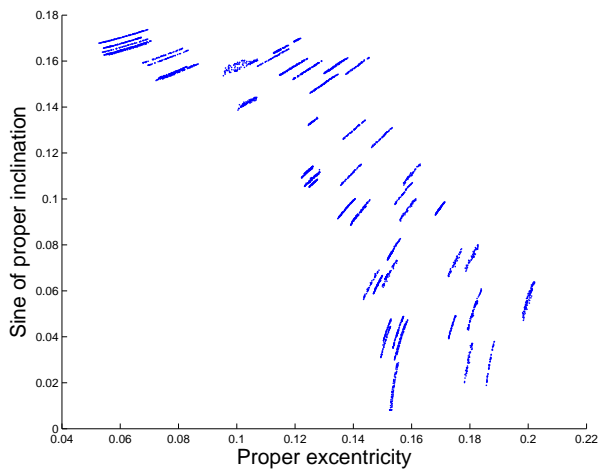}

\caption{A proper $(e,sin{(i)})$ projection of the temporal evolution 
of particles that remained inside the $z_1$ secular resonance for
the whole length of the integration (100 Myr).} 
\label{fig: tracks_ei}
\end{figure}

Vokrouhlick\'{y} {\em et al.} (2006b) introduced the modified
quantity ${K_2}^{'} = \sqrt{1-e^2}(2-\cos{i})$
that is conserved when the sole resonant
gravitational perturbations are taken into account, and that was shown
to be also preserved under the influence of the Yarkorvsky effect 
for timescales of hundreds of Myr.  Similar conserved quantities
can be introduced for other secular resonances as well.
Fig.~\ref{fig: tracks_ei} displays an $(e,sin(i))$ projection of 
the 56 time values of $e$ and $sin(i)$ for the 
76 particles that remained inside the  $z_1 = {\nu}_6+{\nu}_{16}$ 
secular resonance for the whole length of the integration.  It can be
noticed that all particles followed lines of constant ${K_2}^{'}$ and 
oscillated between the maximum and minimum values of $e$ and $sin(i)$ 
allowed by the conservation of ${K_2}^{'}$, as it was also observed
for analogous simulations of resonant particles in the region of the Padua
family by Carruba (2009a).

\begin{figure*}

  \centering
  \begin{minipage}[c]{0.5\textwidth}
    \centering \includegraphics[width=2.5in]{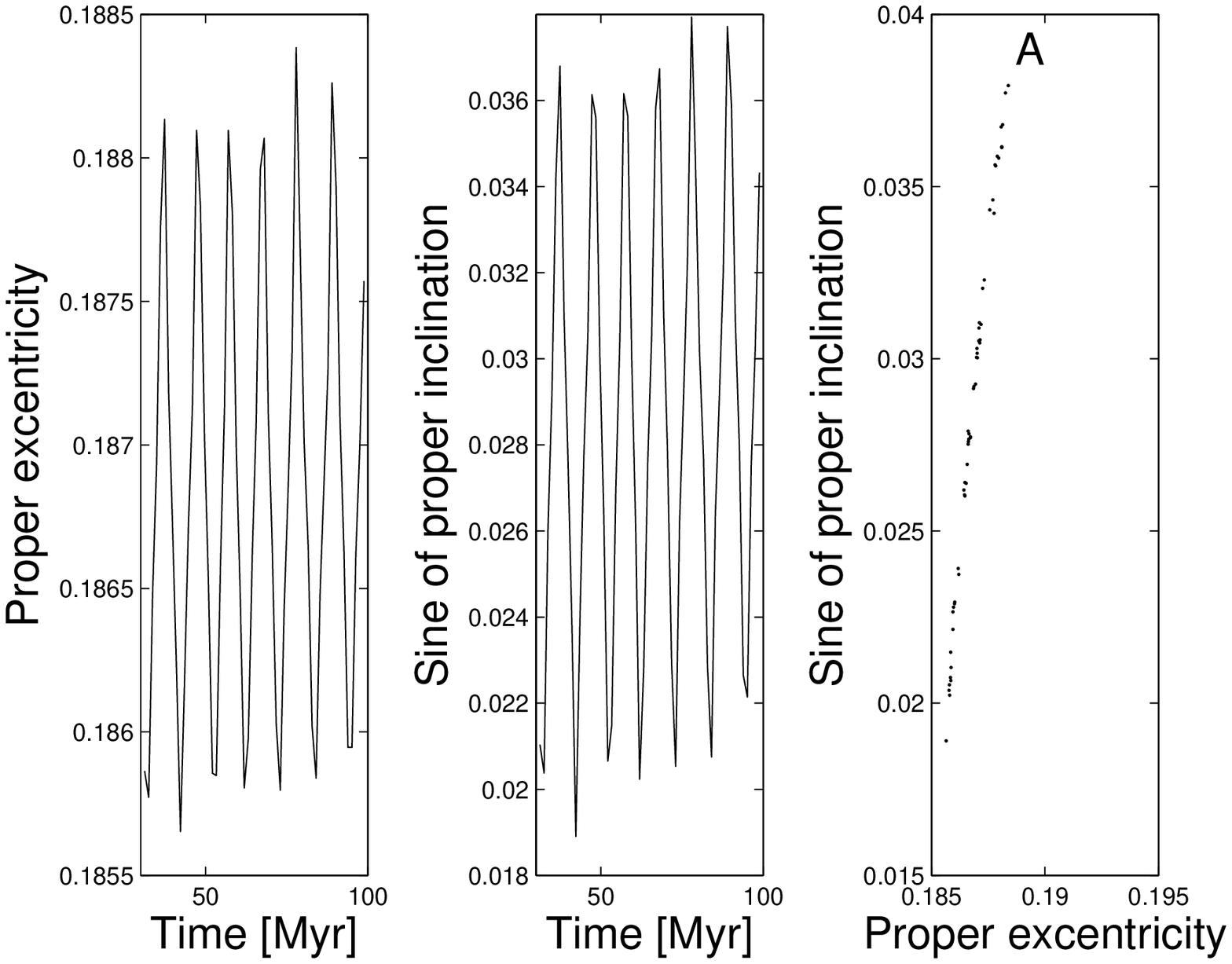}
  \end{minipage}%
  \begin{minipage}[c]{0.5\textwidth}
    \centering \includegraphics[width=2.5in]{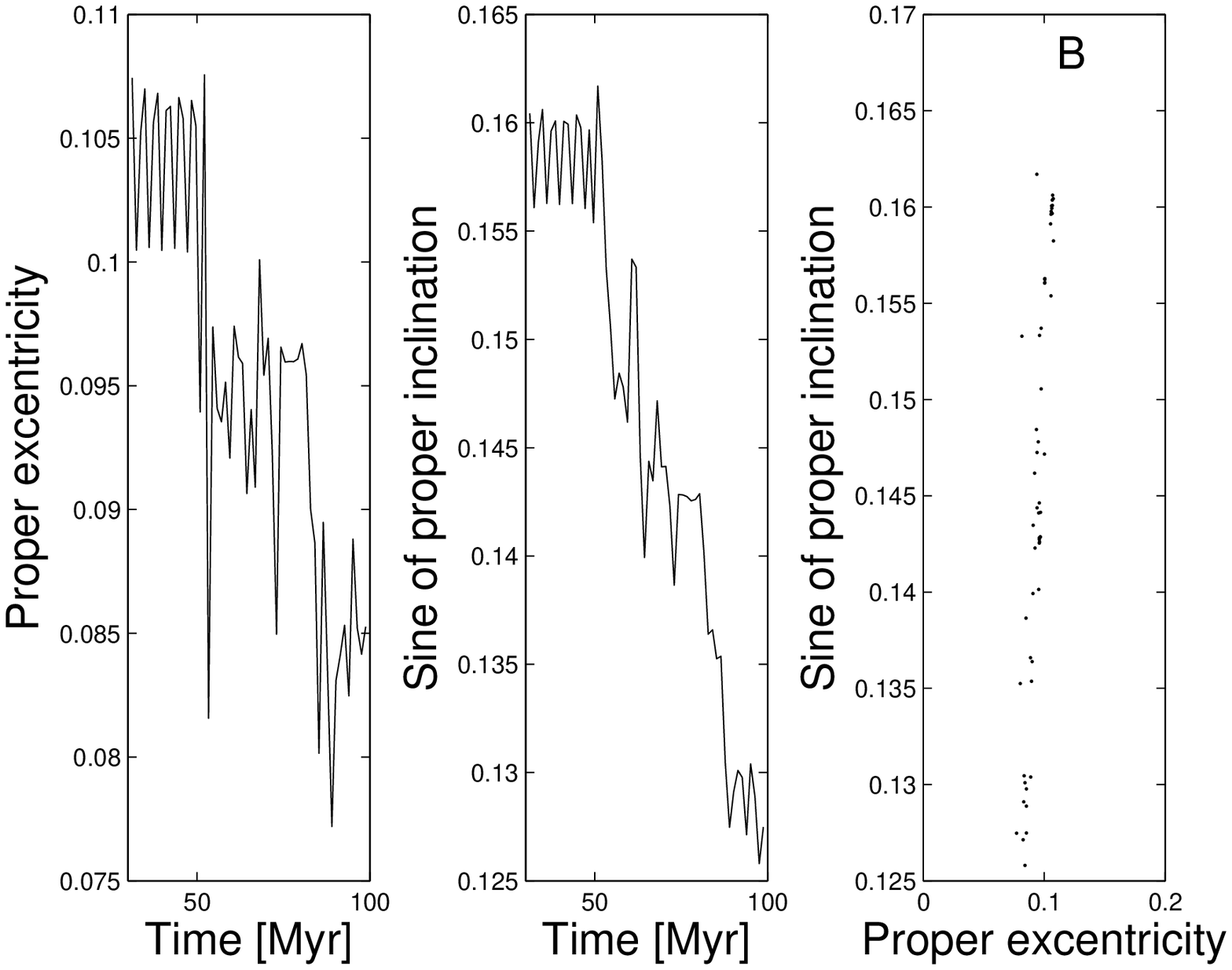}
  \end{minipage}

\caption{The time evolution of proper excentricity,
sine of inclination, and $(e,sin(i))$ of a particle that remained
inside the $z_1$ resonance for the whole length of the integration (panel
A) and another one that escaped around 55 Myr (panel B).
}
\label{fig: plot_esini_3}
\end{figure*}

The scenario is different for particles that escaped the $z_1$ secular 
resonance.  Fig.~\ref{fig: plot_esini_3} displays the time evolution 
of proper excentricity, sine of inclination, and an 
$(e,sin(i))$ projection of a 
particle that remained inside the $z_1$ resonance for the whole 
length of the integration (panel A) and another 
one that escaped after 55 Myr (panel B).
While the first particle displayed the characteristics anti-aligned
oscillations in proper $e$ and $sin(i)$, the particle that escaped
the secular resonance experienced a change in proper $e$ and 
$sin(i)$ at the time of the separatrix crossing of 0.02 and 0.015, 
respectively.  This abrupt and chaotic change caused by the crossing of this 
and several other secular resonances in the region is what may be 
causing part of the asteroid orbital drifting in proper $e$ and $sin(i)$.

\section{Cronology}
\label{sec: cronology}

Having finished our preliminary investigation on mechanisms of dynamical
mobility in the area of the Hygiea family, whose identification was 
discussed in Sect.~\ref{sec: hygiea_det}, we are now ready to start 
addressing the issue of the family age.  We will start by performing
a preliminary analysis based on a simplified model of Yarkovsky drift
(Vokrouhlick\'{y} 1998, 1999).

\subsection{Yarkovsky isolines}
\label{sec: prel_anal}

Vokrouhlick\'{y} {\em et al.} (2006a,b,c) used the $(a,H)$ 
distribution of asteroid families to determine their ages.  In particular,
the authors introduced a parametric target function $C$ defined as:

\begin{equation}
0.2H=log_{10}(\Delta a/C),
\label{eq: target_funct_C}
\end{equation}  

\noindent
where $\Delta a = a-a_c$, and $a_c$ is the ``central'' value of semi-major axis
of the family members.  The authors used distributions of $C$ values
of observed asteroid families members for comparison, using
${\chi}^2$ techniques, with results
of Monte Carlo simulations of diffusion via Yarkovsky and 
Yarkovsky -O'Keefe -Radzievsky -Paddack (YORP) effects to obtain
estimates of family age, and parameters associated with the strength of
the initial ejection velocity field and YORP effect.   
In essence, the YORP effect forces the spin axes of asteroids to 
evolve towards the direction perpendicular to the orbital 
plane.  In this configuration, the semi-major axis drift caused
by the Yarkovsky effect is maximized.  Asteroids either drift 
towards smaller $a$ (if their rotation is retrograde) or larger $a$
(if their rotation is prograde).  This depletes the center of the 
family in the semi-major axis distribution.

The choice of using the absolute magnitude $H$ was justified by 
the limited data then available on asteroids diameters and geometric
albedo values.  For instance, in Vokrouhlick\'{y} et al. (2006a) the 
authors investigated four young families: Erigone, Massalia, Merxia, 
and Astrid.  Only six albedo values were available for the Erigone 
family, and only two for the Merxia group, including an interloper.
New results from the (Wright et al. 2010), and the NEOWISE 
(Mainzer et al. 2011) enhancement 
to the WISE mission allowed to obtain diameters and 
geometric albedo values for more than 100,000 Main Belt asteroids 
(Masiero et al. 2011).  For the case of the Hygiea family, out of the 376 
members of the halo, 280 have WISE values of diameters and geometric 
albedo $p_V$.  Adopting the value of $p_V$ of (10) Hygiea
itself from the WISE mission (that is actually extremely close
to the mean value of geometric albedo of the family, see 
Sect.~\ref{sec: hygiea_det}), Eq.~\ref{eq:  bowell}
can be used to estimate the diameters of the remaining 96 objects.
Since Pravec et al. (2012) have also presented evidence of
possible biases in the various catalogs of asteroid absolute
magnitudes, in this work we choose to modify Eq.~\ref{eq: target_funct_C}
by computing new ``WISE revised'' values of absolute magnitudes $H_{rev}$
with the formula:

\begin{equation}
H_{rev} = 15.617 -5 \cdot log_{10}(2R) -2.5 \cdot log_{10}(p_V),
\label{eq: rev_H}
\end{equation}  

where $R$ is the value of the asteroid radius from the WISE mission.
For asteroids without WISE data on $R$ and $p_V$, we kept the AstDys 
published value of $H$.

A problem in obtaining reliable $C$ values resides in 
the determination of the family
``center''.  The most appropriate definition of family 
center relates to the concept of barycenter (Carruba 2010a,b).  Simulations
of asteroid dynamical groups in the orbital region of the Phocaea family
showed that, while the orbital position of individual asteroids can be 
modified with time by several mechanisms of orbital mobility, the position
of the family barycenter tends to remain relatively stable, and changes
less than the position of the family center.
To compute the barycenter in proper $a$ we took

\begin{equation}
a_c = \sum_{i=1}^{n_{ast}} \frac{a\cdot M_i}{M_{tot}},
\label{eq: bary_a}
\end{equation}

\noindent
where $n_{ast}$ is the number of family members, and 
$M_i$ is the mass of each asteroid, estimated assuming
that all asteroids can be approximated as spheres, using
the density of 2080 $kg/m^3$ reported by Baer et al. (2011) for (10) Hygiea,
and the diameters from the WISE and NEOWISE missions, when available (we 
used the results from Eq.~\ref{eq: bowell} with the geometric
albedo value of (10) Hygiea for the other 96 cases).  Eqs. similar
to Eq.~\ref{eq: bary_a} hold for proper $e$ and $i$.

\begin{table}
\begin{center}
\caption{{\bf Location in proper elements $(a,e,sin(i))$ domain
of the barycenter of (10) Hygiea, the Hygiea family core and halo}}
\label{table: hygiea_bar}
\vspace{0.5cm}
\begin{tabular}{|c|c|c|c|}
\hline
         &                &                 &                        \\
Group    & $a_{barycenter}$  &  $e_{barycenter}$  &  ${sin(i)}_{barycenter}$  \\
         &                &                 &                        \\
\hline
         &                &                 &                        \\
(10) Hygiea        & 3.1262 & 0.1493 & 0.0895                        \\ 
Hygiea family core & 3.1390 & 0.1346 & 0.9144                        \\
Hygiea family halo & 3.1423 & 0.1224 & 0.0953                        \\
         &                &                 &                        \\
\hline
\end{tabular}
\end{center}
\end{table}

We computed the location of the barycenter of the family core and halo, with 
and without (10) Hygiea, which holds more than 95\% of the mass of 
the family.  Expectedly, if we include (10) Hygiea, the position of the 
family barycenter differs little from that of (10) Hygiea itself.  
Nevertheless, if we exclude this asteroid and other objects 
unlikely to be members for
dynamical reasons that will be explained later on in this section, there
is a discrepancy in the position of the barycenter of $0.016~AU$
with respect to the current location of (10) Hygiea.   Since the family
shows an asymmetry, with more objects at larger $a$ than at lower $a$,
if we put the center of the family at the current orbital location of 
(10) Hygiea, we believe that a possible reason for this discrepancy could be 
that the current position of (10) Hygiea is not the one that this asteroid
occupied when the family formed.  While the Yarkovsky drift for a body this
size ($D = 453.2~km$, Masiero et al. 2011) is practically infinitesimal
even on timescales of the order of the age of the Solar System, 
in paper II it was observed that (10) Hygiea experienced close 
encounters with (1) Ceres, (2) Pallas, (4) Vesta and other 
massive asteroids that could account for a displacement in proper $a$ of
this asteroid of the observed amount.  The discrepancy between the current 
position of the barycenter of the family (without (10) Hygiea) 
and the current position of (10) Hygiea itself could be a ``fossil'' proof
of past dynamical mobility of the orbit of this asteroid caused by 
close encounters with massive asteroids~\footnote{Other
possible explanation of this asymmetry include an oblique
collision between the impactor and the parent body of the Hygiea family,
or the possible existance of multiple Hygiea families, caused by 
successive collisions.  In this work we prefer to investigate 
what, in our opinion, is the simpler
scenario, that involves a non atypical ejection velocity field, 
a single Hygiea family, and a limited amount of dynamical mobility
for (10) Hygiea itself.}.
Table~\ref{table: hygiea_bar} summarizes the results of our 
analysis, with values of the barycenter location in proper $a, e$ and $sin(i)$
for (10) Hygiea itself, and the barycenter of the family (without (10) Hygiea
and other dynamical interlopers) core and halo, as obtained in 
Sect.~\ref{sec: hygiea_det}.  Almost all the mass of the family 
is concentrated in (10) Hygiea, with only 0.57\% of the total mass 
in halo members, and 0.34\% in core members.

\begin{figure*}

  \centering
  \begin{minipage}[c]{0.5\textwidth}
    \centering \includegraphics[width=2.5in]{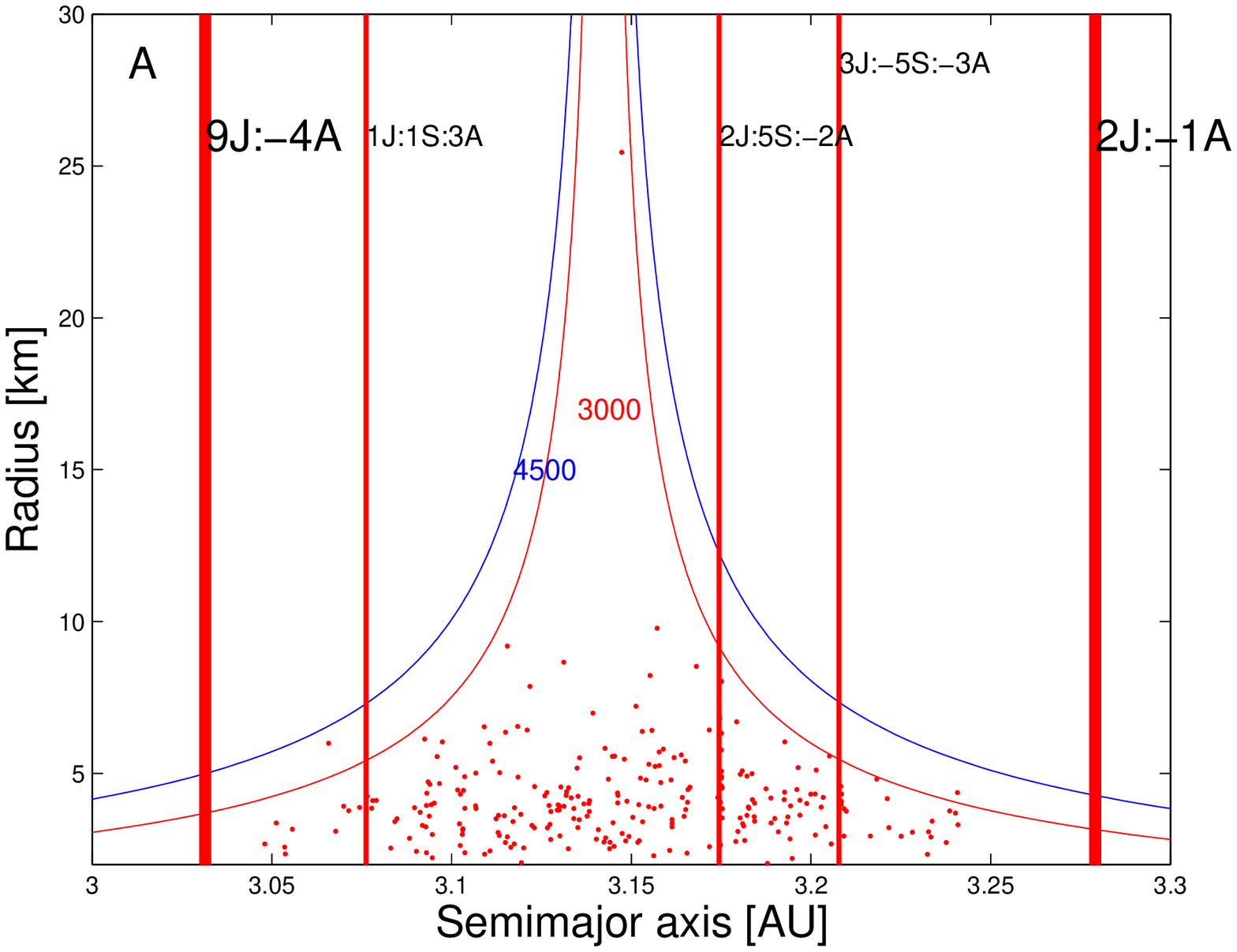}
  \end{minipage}%
  \begin{minipage}[c]{0.5\textwidth}
    \centering \includegraphics[width=2.5in]{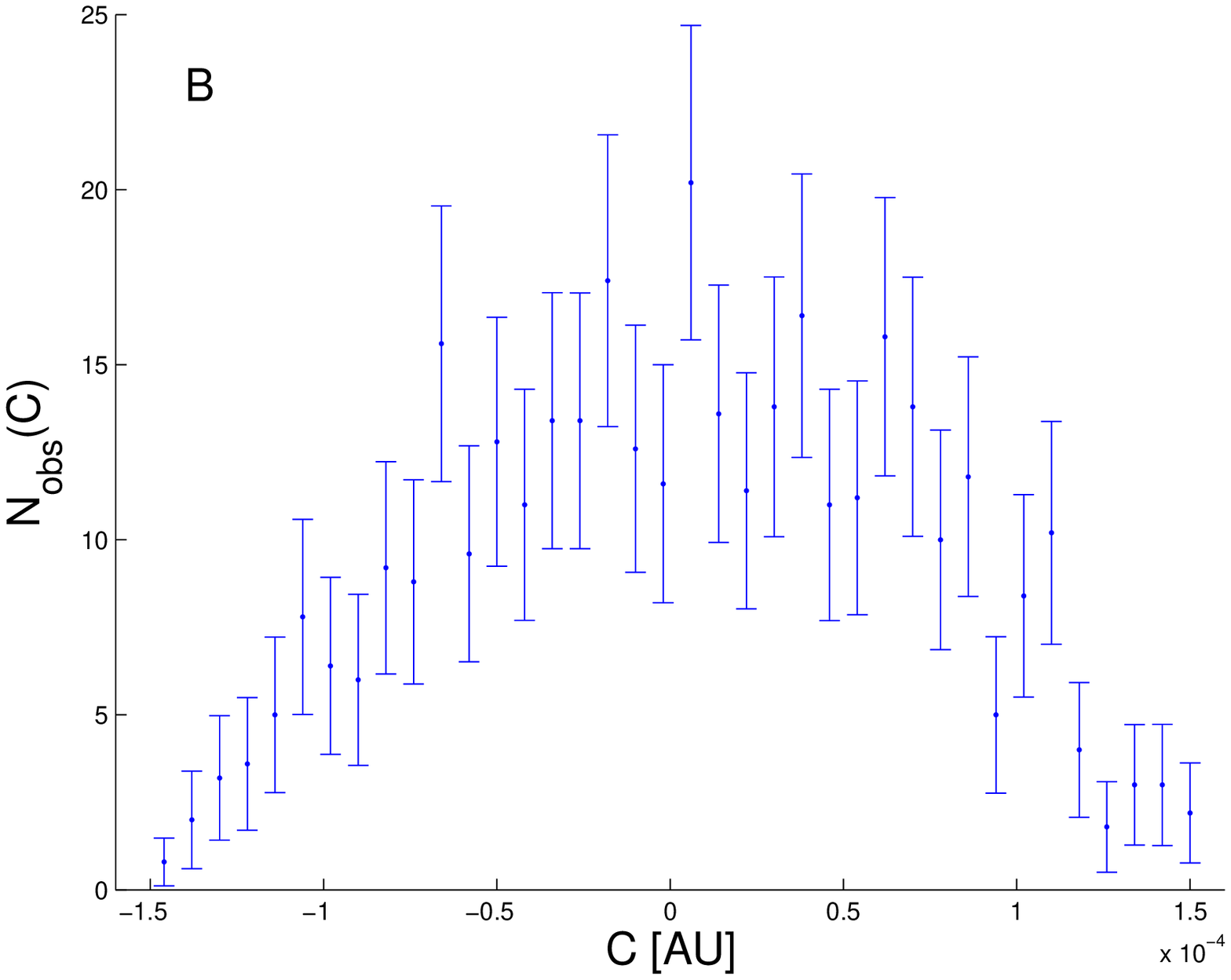}
  \end{minipage}

\caption{Panel A: a proper $a$ versus radius projection of 
members of the Hygiea family halo.  The red and yellow curves
show isolines of maximum displacement in $a$ caused by Yarkovsky
effect for a fictitious family originally centered in the family
barycenter after 3.0~Byr (red line) and 4.5~Byr (blue line).
Panel B:  values of the $C-$target function 
for the same objects.  The error bars are given as $\sqrt{N_{obs}(C)}$ for
each interval $(C,C+\Delta C)$, with $\Delta C = 8.0 \cdot 10^{-6}$~AU.
}
\label{fig: V_yarko}
\end{figure*}

\begin{figure}

  \centering
  \centering \includegraphics [width=0.45\textwidth]{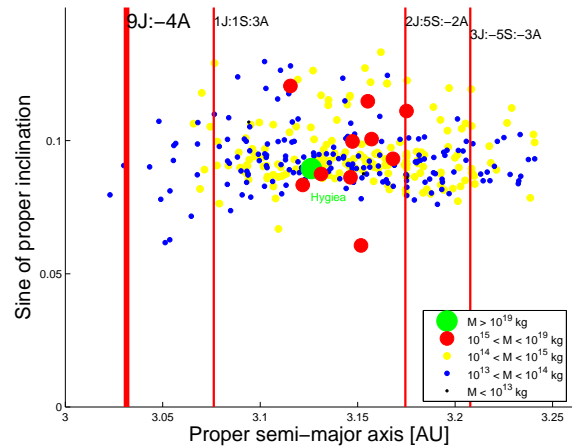}

\caption{An $(a,sin(i))$ projection of the Hygiea halo members.  The 
sizes of the dots are proporcional to the asteroid mass.} 
\label{fig: mass_distr}
\end{figure}

We then turned our attention to the Yarkovsky isolines method to 
obtain asteroid age estimates and the $C$-parameter distribution.
Fig.~\ref{fig: V_yarko}, panel A, displays a proper $a$ versus 
radius $R$ projection of members of the Hygiea family halo (results are 
similar in terms of age determination for Hygiea family core
members and won't be shown for simplicity; also, with a radius
of $226.6~km$ (10) Hygiea itself is out of the range of this 
picture).  The red and 
blue curves show isolines of maximum displacement in $a$ caused by the 
Yarkovsky effect, computed using the Vokrouhlick\'{y} (1998, 1999) model of 
the diurnal version of the Yarkovsky effect, for spherical bodies and 
in the linear approximation for the heat conduction in a spherical, 
solid, and rotating body illuminated by solar radiation, 
for a fictitious family originally centered in the family
barycenter after 3.0 Byr (red line) and 4.4 Byr (blue line).  We 
used the following parameters to describe the Yarkovsky force:
a value of thermal conductivity $K = 0.001 $W/m/K,  a specific 
heat capacity of $C_p = 680$ J/kg/K, a density 
of 2080 $kg/m^3$, a surface density of 1500 $kg/m^3$, a bond albedo
of 0.11, the geometric albedo from the WISE mission 
when available (otherwise
we used the value of geometric albedo of (10) Hygiea), 
and the diameters of halo and core members previously computed.
We eliminated all C-, B-, and X-type asteroids
whose distance from the family barycenter was larger than the maximum 
Yarkovsky drift over 4.5 Byr plus $0.02~AU$, the maximum displacement
caused by close encounters with (10) Hygiea observed in paper II.  
Asteroids (52) Europa, (106) Dione, (159) Aemilia, 
(211) Isolda, (538) Friederike, (867) Kovacia, 
(1107) Lictoria, (1599) Giomus, (2436) Hatshepsut, 
(6644) Jugaku, (9544) Scottbirney, (16093) (1999 TQ180), and 
(16450) Messerschmidt were
considered dynamical interlopers, and their
mass was not considered in the computation of the mass of the family 
and of the family barycenter.  After this analysis, we were left
with a sample of 267 and 363 possible Hygiea family members in the 
core and in the halo, respectively.  Fig~\ref{fig: mass_distr} 
displays an $(a,sin(i))$ projection of the halo members,
with the sizes of the dots displayed according to the asteroid
mass.  As previously discussed, more than 99\% of the family mass
is owned by (10) Hygiea itself.

Yarkovsky isolines will not give a very reliable estimate of the age 
of the family, since they do not account for the original dispersion
at break-up moment (all asteroids are assumed to be originally at the
barycenter of the family) and do not consider the effect of reorientations
and YORP cycles (Vokrouhlick\'{y} and \v{C}apek 2002,
\v{C}apek and Vokrouhlick\'{y} 2004), but they will give a 
preliminary value that can later
on be refined by more advanced method, as the ones discussed in the 
next sections.  The Hygiea family seems to be a relatively old group, with 
an age at least larger than $3.0~Byr$.  Since modeling of YORP cycle and
effect of reorientations are not reliable for ages larger than
$\simeq 1.0~Byr$ (Vokrouhlick\'{y} {\em et al.} 2006a,b), a very 
accurate determination of the family age seems to
be not likely to occur.  Nevertheless, we will try in the next sub-section
to further refine this initial order of magnitude estimation.

Fig.~\ref{fig: V_yarko}, panel B shows a histogram of the values of 
the $C$ target function for the Hygiea family halo members computed with 
Eqs.~\ref{eq: target_funct_C} and ~\ref{eq: rev_H}, after our process of 
elimination of taxonomical and dynamical interlopers.  We used 38
intervals starting at $C_{min} = -1.5 \cdot 10^{-4}$ with a step of 
$8.0 \cdot 10^{-6}$~AU.  We computed an average
$C$ distribution for the halo family obtained as a mean of the 
computed values for $a_c$ in the interval $[3.140,3.144]~AU$ around
the family barycenter (computed without (10) Hygiea and
dynamical interlopers).  This was done to avoid random fluctuations
in the $N(C)$ distribution that would affect the quality of the fit.
The Hygiea family seems to show a not very pronounced
bimodal distribution of $C$ values, typical of Yarkovsky/YORP 
evolved asteroid families.
Synthetic distributions of the $C$ function
will be obtained using Monte Carlo simulations of Yarkovsky and YORP
dynamical mobility of fictitious family members in the next sub-section.

\subsection{Monte Carlo simulations}
\label{sec: montecarlo}

\begin{figure}

  \centering
  \centering \includegraphics [width=0.45\textwidth]{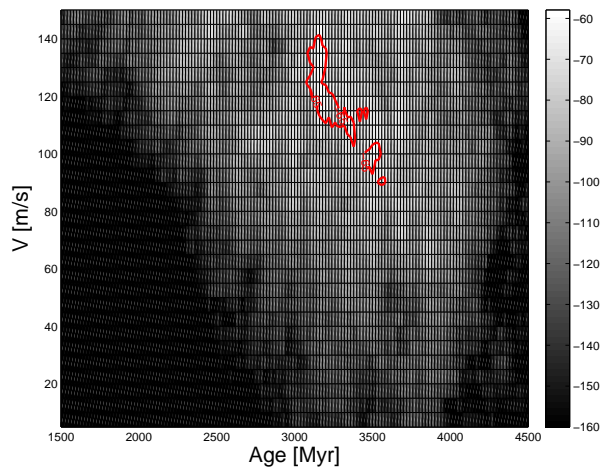}

\caption{Values of the target function 
${\psi}_{\Delta C}$ in the ($Age,C_{YORP}$) plane for member of the 
Hygiea family halo.} 
\label{fig: AGE_V_halo}
\end{figure}

In this section we will analyze the semi-major axis evolution
of the Hygiea halo family members.  The methods used here follow the
work of Vokrouhlick\'{y} {\em et al.} (2006a, b), that showed that
the skewed $a$-distribution of asteroid family members may be 
explained as a consequence of the Yarkovsky -O'Keefe -Radzievsky -Paddack
(YORP) effect.  Asteroids have their spin axes evolved to alignments
perpendicular to the orbital plane, which accelerates the migration
of asteroids and depletes the family center. Modeling the evolution 
of asteroid members as a function of the parameters that characterize the
Yarkovsky and YORP evolution of the asteroids (age 
of the family, YORP strength ($C_{YORP}$), the characteristic ejection
velocity $V_{EJ}$ given by 

\begin{equation}
V_{SD}=V_{EJ}\cdot(5km/D),
\label{eq: V_EJ}
\end{equation}

\noindent
where $V_{EJ}$ is a free parameter characterizing the size of the family
in velocity space and $D$ is the asteroid diameter, may produce a distribution
of semimajor axis values, that can then be transformed into a $C$-distribution
using Eq.~\ref{eq: target_funct_C}. Using the values of 
Yarkovsky parameters discussed
in the previous subsections, the simulated $C$-distributions 
can then be compared to the observed one, and minimizing the 
${\chi}^2$-like function:

\begin{equation}
{\psi}_{\Delta C}=\sum_{\Delta C}\frac{[N(C)-N_{obs}(C)]^2}{N_{obs}(C)},
\label{eq: psi}
\end{equation}

the best-fit values of the Yarkovsky and YORP model can be obtained.   Since
the predicted age of the Hygiea family is higher than $1.0~Byr$, a value for
which the current understanding of the YORP cycle is not very accurate, 
and since past experiences with this method showed us that is not very
dependent on the value of the $C_{YORP}$ parameter, provided is not zero 
(Carruba 2009a, Masiero et al. 2012), we will assume a value of 
$C_{YORP} = 1.0$, and will limit our analysis
to a two-dimensional parameter space defined by the age of the family 
and its characteristic ejection velocity $V_{EJ}$.
Admissible solutions are characterized by ${\psi}_{\Delta C}$ of the
order of the number of used bins in C (38, in our case), while solutions
giving much larger ${\psi}_{\Delta C}$ are incompatible with the 
observed family.  To quantify the goodness of the fit, we used the 
incomplete gamma function $\Gamma (N_{int},{\psi}_{\Delta C})$ 
(Press {\em et al.} 2001), where $N_{int} = 38 $ is the number of 
intervals used for the values of the $C$ target function and 
${\psi}_{\Delta C}$ was computed with Eq.~\ref{eq: psi}.  We used a value of 
${\psi}_{\Delta C}$ of 65 as a limit for an acceptable fit (red line 
in Fig.~\ref{fig: AGE_V_halo}, as this would correspond to a probability of 
99.67\% that the simulated distribution differs from the observed (Press 
et al. 2001).

Fig.~\ref{fig: AGE_V_halo} shows the values of the target
function ${\psi}_{\Delta C}$ in the ($Age, V$) 
plane~\footnote{To associate lower levels of ${\psi}_{\Delta C}$ 
with whither tones we plotted color plots of $-{\psi}_{\Delta C}$.}. 
The predicted age and the characteristic ejection velocity 
field $V_{EJ}$ of members of the Hygiea family halo are in
the range $T = 3200^{+380}_{-120}$~My and 
$V_{EJ} = 115\pm 26$~m/s, and results are similar for the
Hygiea family core.  Estimates of the Hygiea family age
obtained with Monte Carlo simulation are in good agreement with
results obtained with the simpler approach described in 
\ref{sec: prel_anal}, and with those reported by Nesvorn\'{y} 
et al. (2005).   The relatively 
high value of the ejection velocity parameter,
unusual for families associated with the break-up of the parent body,
is however typical of families created by cratering events, such as
the Vesta and the Hygiea groups.

\subsection{Effects of close encounters on Hygiea family chronology}
\label{sec: ce_chron}

As a next step in our analysis of the dynamical evolution of the Hygiea
family, we investigated the effects that close encounters with (10) Hygiea
may have on the estimate of the family age.  For this purpose we
first best-fitted polynomials of order eight to the $pdf$ 
distribution of changes in proper $a$ (as obtained in Sect.~\ref{sec: pdf})
between the values $0.0006 < |\Delta a| < 0.006$~AU, and
a gaussian with the measured values of standard deviation in 
proper $a$ of $8.9\cdot 10^{-4}~$AU for the central values.
We then replicated the observed distribution using the 
rejection method (Press et al. 2001), with appropriate Lorentzian
distributions as comparison probabilities distribution functions.

\begin{figure}

  \centering
  \centering \includegraphics [width=0.45\textwidth]{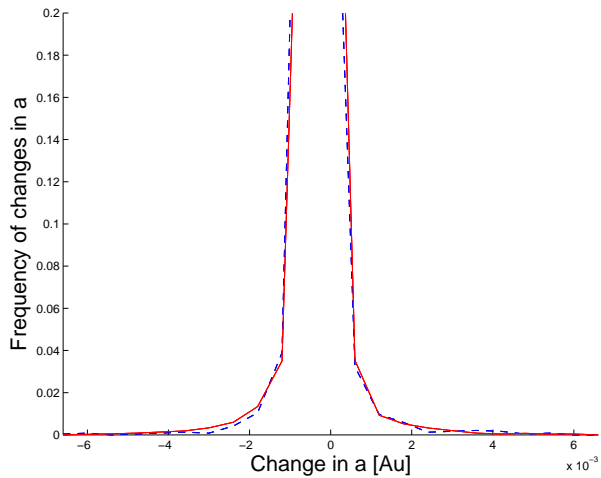}
\caption{A histogram of the $pdf$ curve
obtained in Sect.~\ref{sec: pdf} (red curve) and of of the new probability
distribution function obtained with the rejection method (blue dotted
line).} 
\label{fig: lorentz_rej}
\end{figure}

Fig.~\ref{fig: lorentz_rej} displays a histogram of the $pdf$ curve
obtained in Sect.~\ref{sec: pdf} (red curve) and of of the new probability
distribution function obtained with the rejection method (blue dotted
line), computed over 23 equally spaced intervals between $\pm 0.0066$~AU.
Percentual differences between values of the two distributions are less than 
1\%.  We then performed new Monte Carlo simulations of the dynamical evolution
of synthetic Hygiea family members, that include this time the effect 
of close encounters with (10) Hygiea.  We computed changes in semi-major
axis caused by the computed $pdf$ after each 120.2 Myr, time sufficient for
having a $2\sigma$-compatible $fdf$ and a population of close encounters
equal to seven times the number of Hygiea halo members, by applying
seven times the computed $pdf$, and obtaining a resulting cumulative change
in proper $a$ for each of the 363 members of the Hygiea halo.

\begin{figure}

  \centering
  \centering \includegraphics [width=0.45\textwidth]{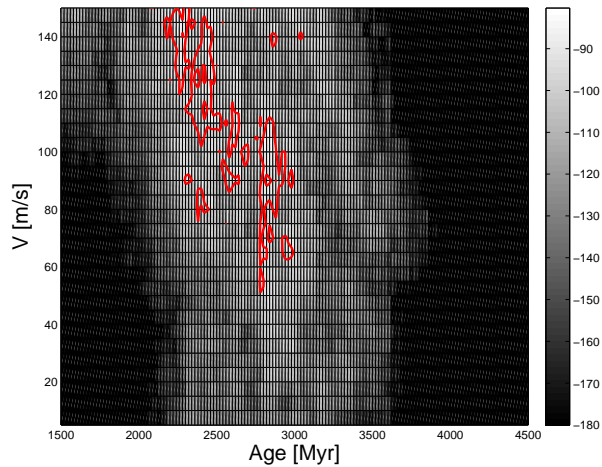}

\caption{Values of the target function 
${\psi}_{\Delta C}$ in the ($Age,C_{YORP}$) plane for member of the 
Hygiea family halo, evolved under the influence or Yarkovsky and 
YORP effects, and close encounters with (10) Hygiea.} 
\label{fig: AGE_V_halo_CE}
\end{figure}

Fig.~\ref{fig: AGE_V_halo_CE} shows the values of the target
function ${\psi}_{\Delta C}$ in the ($Age, V$) plane for this
new set of simulations.  We used a value of 
${\psi}_{\Delta C}$ of 93 as a limit for an acceptable fit (red line 
in Fig.~\ref{fig: AGE_V_halo_CE}. The predicted age and the characteristic 
ejection velocity field $V_{EJ}$ of members of the Hygiea family halo 
are now in the range $T = 2420^{+580}_{-270}$~My and 
$V_{EJ} = 135^{+15}_{-85}$~m/s, and results are similar for the
Hygiea family core.  Surprisingly, we found that, with respect
to the simulations without close encounters with (10) Hygiea,
the estimated age of the family is 780 Myr younger, which corresponds
to a change of 24.4\%.   This new result poses new interesting
questions on the importance of close encouters for the dynamical evolution of
asteroid families around massive bodies, such as the Hygiea, Vesta,
Gefion, and, possibly, the Pallas family.

\section{Conclusions}
\label{sec: concl}

In this work we:

\begin{itemize}

\item Performed a taxonomical analysis of members of the Hygiea
family (core and halo) as obtained in Carruba (2013), with the 
DeMeo and Carry (2013) approach.  We identified 276 and 376
members of the Hygiea family core and halo, respectively, 
whose taxonomy is compatible with membership in this family (types
C-, X-, and B-), and have $H < 13.5$.  
Except for B-type objects, that may possibly be
associated with the Hygiea and Themis family, C- and X-types bodies
in the Hygiea halo cannot univocally be associated with the break-up
of the Hygiea parent body, but can possibly also be originating from
the Veritas or Themis families.

\item Studied the long-term effect of close encounters with (10) Hygiea
on the dynamical evolution of members of its halo.  We confirmed
the prediction of Carruba et al. (2013) that the frequency distribution
function ($fdf$) of changes in $a$ converges to the probability distribution
function ($pdf$) for number of encounters of about 6000 (5455). The 
populations of Hygiea family halo and core identified with the 
DeMeo and Carry (2013) approach are expected to experience a number of
encounters sufficient to obtain a $2\sigma$ level (or 95.4\%) approximation
of the $pdf$ in 113.1 and 162.3 Myr, respectively.

\item Studied the effect of secular dynamics in the region of the Hygiea 
asteroid family.  The secular resonances that cross the dynamical
family, such as the ${\nu}_5+2{\nu}_{16}$,
${\nu}_6+2{\nu}_{16}$, $2{\nu}_6-{\nu}_5+{\nu}_{16}$, and 
$2{\nu}_6-{\nu}_5+{\nu}_{17}$ resonances, have a limited current 
population of resonators and relatively short sticking times.
Furthermore, they influence the orbital evolution of Hygiea family 
members mostly by the occasional relatively large change in proper 
elements caused by the crossing of the resonance separatrix.
Of the three most populated secular resonances studied in Carruba 2013,
the $3{\nu}_6-2{\nu}_5$, $2{\nu}_5-2{\nu}_6+{\nu}_{16}$, and ${\nu}_6+{\nu}_{16}$
resonances, the $z_1 = {\nu}_6+{\nu}_{16}$ resonance has the largest current
population of resonators and the longest sticking times. 

\item Obtained a preliminary estimate of the age of the Hygiea family based on 
the method of Yarkovsky isolines of at least 3.0 Byr.  
We found that (10) Hygiea itself is not 
located at the current position of the barycenter of the family, but 
it is displaced by $0.016~AU$.  This may indicate that it experienced 
dynamical mobility since the formation of the family, 
possibly caused by close encounters with some other massive asteroids 
since Yarkovsky mobility is negligible for a body of its size.

\item We performed Monte Carlo simulations of the dynamical mobility 
caused by the Yarkovsky and YORP effects of synthetic Hygiea family members
following the approach of Vokrouhlick\'{y} {\em et al.} (2006a, b).  
The predicted age and the characteristic ejection velocity 
field $V_{EJ}$ of members of the Hygiea family halo are in
the range $T = 3200^{+380}_{-120}$~Myr and $V_{EJ} = 115\pm 26$~m/s, and 
results are similar for the Hygiea family core, and in agreement 
with previous estimates (Nesvorn\'{y} et al. 2005).

\item We modelized the long-term effect of close encounters on the dynamical
evolution in semi-major axis of members of the Hygiea family halo.
Surprisingly, we found that including close encounters as a mechanism
of dynamical mobility could reduce the estimated age of the Hygiea asteroid
family by $\simeq$ 25\%.  This poses new interesting
questions on the importance of close encouters on the dynamical evolution of
asteroid families around massive bodies, such as the Hygiea, Vesta,
Gefion, and, possibly, the Pallas family.

\end{itemize}

As a by-product of our taxonomical analysis of the Hygiea family halo,  
we identified, very surprisingly, one possible V-type candidate: 
the asteroid (177904) (2005 SV5).  If confirmed, this
could be the fourth V-type object ever to be identified in the outer main 
belt.  

While in this work we obtained the first estimate of the Hygiea family
age that we are aware of, we would like to emphasize that this 
estimate is affected by several uncertainties: Masiero et al. (2012)
showed that changes in the values of thermal conductivities and mean
densities of family members can change the estimated value of
the family age by up to 40\%.  Other effects such the progressive
change in surface properties caused by space weathering, the 
change in solar luminosity in the past (Bahcall et al. 2001), and,
possibly, low energy collisions (Dell'Oro and Cellino 2007) all may 
play a role in affecting the estimated age of the family.  
While we do not yet have a good undestanding of how to model the space
weathering effect on C-, B-, and X-type asteroids, the effect
of changes in solar luminosity affects the estimated age of the family
by at most 4\% (Vokrouhlick\'{y} et al. 2006b), and the effect 
of low-energy collisions seems to be a minor one (Carruba 2009a),
of most importance however, are the current limitations on our understanding
of the YORP effect.  Cotto-Figueroa et al. (2013) have shown that the 
YORP effect has an extreme sensitivity to the topography of asteroids. 
If the spin-driven reconfiguration leads to a shape of the aggregate 
that is nearly symmetric, the YORP torques could become negligibly small 
or even vanish. This would imply a self-limitation in the evolution of 
the spin state and the objects would not follow the classical YORP cycle. 
Since our understanding of YORP cycles may be not adequate for
modeling the evolution of family members on very long time scales, 
we preferred not to conduct extensive simulations with symplectic 
integrators in order to better refine the age estimate of the Hygiea 
family.

Whatever the best way to account for all the limitations on our modeling
of the long-term effects of Yarkovsky and YORP effect could be, here for
the first time we showed the importance that close encounters with massive
asteroids may also have for the dynamical evolution of the Hygiea family, 
and how they could reduce the estimated age of this family by $\simeq$ 25\%.
This opens new and interesting perspectives for investigating the ageing
process of asteroid families, that could be best investigated by 
studying younger families with relatively massive parent bodies, such as,
possibly, the Massalia family (Vokrouhlick\'{y} et al. 2006a).

\section*{Acknowledgments}
We thank the reviewer of this paper, Ricardo Gil-Hutton, for 
comments and suggestions that significantly increased the quality of
this work.  We also would like to thank the S\~{a}o Paulo State 
Science Foundation (FAPESP) that supported this work via the grant 
11/19863-3, and the Brazilian National Research Council (CNPq, grant 
305453/2011-4). This publication makes use of data products from the 
Wide-field Infrared Survey Explorer, which is a joint project of the 
University of California, Los Angeles, and the Jet Propulsion 
Laboratory/California Institute of Technology, funded by the National 
Aeronautics and Space Administration.  This publication also makes use of 
data products from NEOWISE, which is a project of the Jet Propulsion 
Laboratory/California Institute of Technology, funded by the Planetary 
Science Division of the National Aeronautics and Space Administration.

\bsp

\label{lastpage}

\end{document}